# A Novel Reinforcement Learning Model for Post-Incident Malware Investigations

Dipo Dunsin, Mohamed Chahine Ghanem *, Karim Ouazzane, Vassil Vassilev,

***Abstract*—This Research proposes a Novel Reinforcement Learning (RL) model to optimise malware forensics investigation during cyber incident response. It aims to improve forensic investigation efficiency by reducing false negatives and adapting current practices to evolving malware signatures. The proposed RL System leverages techniques such as Q-learning and the Markov Decision Process (MDP) to train the system to identify malware patterns in live memory dumps, thereby automating forensic tasks. The RL model is based on a detailed malware workflow diagram that guides the analysis of malware artefacts using static and behavioural techniques as well as machine learning algorithms. Furthermore, it seeks to address challenges in the UK justice system by ensuring the accuracy of forensic evidence. We conduct testing and evaluation in controlled environments, using datasets created with Windows operating systems to simulate malware infections. The experimental results demonstrate that RL improves malware detection rates compared to conventional methods, with the RL model's performance varying depending on the complexity and learning rate of the environment.**

## I. Introduction

In post-incident malware forensics investigations, the detection and classification of malware are critical processes for reconstructing evidence files. This is particularly important because malware, being a malicious program, can lead to unauthorised access to confidential information, jeopardising the security and integrity of data or information systems, thereby posing a significant threat to the involved systems and institutions. He and Sayadi [1] highlight that malware attacks have become a pervasive threat, affecting homes, education, businesses, government, and healthcare by "finding vulnerabilities in networks and applications to launch attacks." In the healthcare field, particularly within the Internet of Medical Things (IoMT), such malicious attacks are especially dangerous [2]. A minor misclassification or failure to detect malware can seriously compromise patient medical records, potentially leading to incorrect diagnoses or treatments and, in extreme cases, resulting in paralysis or death. Machine Learning (ML) has gained widespread adoption to detect various types of malware. However, increasingly complex malware can circumvent ML techniques and models. The paper by Wu et al., [4] discusses the use of reinforcement learning (RL) as a more advanced model that improves malware detection accuracy, surpassing traditional machine learning methods. According to Liu et al., [5], machine learning in malware detection involves extracting features from data to classify malware and learn from past data to identify new threats [6] [7]. Ebrahimi et al., [8] suggest that reinforcement learning can help defenders detect sophisticated adversarial threats. Adversarial malware relies on perturbation methods to evade ML-based detectors [9] [10]. A significant challenge is the need for constant updates with new malware behaviours.

Reinforcement learning, on the other hand, enables models to generate adversarial malware that can bypass detection by portable executable (PE) malware classifiers. Quertier et al., [10], note that the RL System "Gym-Malware" achieves an evasion rate of up to 16%, while the RL System "MAB-malware" can achieve an evasion rate of over 75% in a black-box setting. Furthermore, according to Quertier et al., [10], adversarial knowledge defines attacks as either "white box," where the adversary has complete access to the model, or "black box," where the adversary has no knowledge of the model and can only obtain classification results through a limited number of attempts [13]. These adversarial samples, which are falsely classified as benign, can improve the detection accuracy of malware detectors by 16% to 94% [4]. Reinforcement learning's ability to automate malware evasion shows how detailed sequences of adversarial actions can train antivirus and malware detection systems, enhancing their effectiveness in combating malware [10].

### A. Research Aim and Question

This research aims to improve malware forensics investigations by utilising reinforcement learning (RL) techniques. The primary focus is on identifying, analysing, and enhancing models for post-incident investigations. As a result of this, the goal is to expedite forensic processes and mitigate the miscarriage of justice within the UK legal system [3]. Additionally, this research seeks to improve heuristic- and signature-based analysis methods through the application of RL, thereby enhancing overall cybersecurity measures after a security breach. Building on these objectives, the research central question is: How effective are reinforcement learning models in distinguishing between benign and malicious software, and what are the areas for potential improvement ?

* Mohamed C. Ghanem is the corresponding author. email: m.ghanem@londonmet.ac.uk| Mr. D. Dunsin, Dr. M.C. Ghanem, Prof. K. Ouazzane and Prof. V. Vassilev are with the Cyber Security Research Centre, London Metropolitan University, London, UK

## II. Related work

### A. Reinforcement Learning Improves ML Malware Detection

Wu et al. [4] explore reinforcement learning (RL) to enhance malware evasion against machine learning detection models. They propose the "gym-plus" model, an extension of "gym-malware," which generates evasive malware samples. Using the EMBER dataset, they retrain detection models, improving detection rates for unknown malware. However, the study lacks a clear theoretical explanation of RL and omits a discussion on limitations like agent selection and action space. Moreover, there is no comparative analysis with other detection methods. Despite these gaps, the paper demonstrates practical effectiveness, offering a detailed evaluation of the RL model [21].

### B. Adversarial RL for Malware Detection

Ebrahimi et al., [8] research improves cybersecurity by addressing adversarial attacks on ML-based malware detection. Specifically, their innovative use of adversarial reinforcement learning (RL) enhances malware detector robustness in dynamic environments. The method improves dynamic interactions between the detector and the adversary by including adversarial agents that create false samples. However, reliance on MalVAC and lack of comparative assessment limit the research's generalizability.

### C. Reinforcement Learning: Uncovering Control

Wang et al., [11] propose an automated approach for identifying command and control (C2) attack paths in large networks using reinforcement learning (RL). They address manual C2 detection methods' limitations by emphasizing efficiency and automation. By integrating cyberdefense terrain into the RL model, they enhance practical relevance. While their detailed attack simulation, which covers infection to exfiltration phases, demonstrates rigor, reliance on a simulated environment limits real-world applicability. Furthermore, the assumption of full knowledge of host characteristics may be unrealistic, and the RL model's complexity poses practical challenges. D

### D. Reinforcement Learning for Grid Security

Yu [12] introduces a hierarchical deep reinforcement learning-based (HDRAD) scheme for detecting Advanced Persistent Threats (APT) in data management systems. The thorough simulations demonstrating HDRAD's superior performance over RS and HP schemes in detection delay and data protection are a key strength. However, the study's complexity and reliance on technical illustrations make it difficult for non-experts. While the research outlines its objectives, it lacks simplified explanations of how the HDRAD scheme functions.

### E. Deep Learning Techniques for Malware Obfuscation

Gao and Fang [14] present a novel malware evasion approach using deep reinforcement learning. They generate adversarial examples by extracting bytes from benign files and injecting them into malware, achieving an 85% evasion rate against EMBER, a state-of-the-art malware classifier. The research's strengths include the innovative combination of reinforcement learning and malware evasion, along with a clear presentation. However, weaknesses arise from a lack of discussion on ethical implications and potential limitations, such as effectiveness across diverse malware types.

### F. A3C Algorithm for Malware Identification

Xue et al., [16] describe a novel approach to improving malicious code detection using reinforcement learning and the Asynchronous Advantage Actor-Critic (A3C) algorithm. They highlight the limitations of traditional detection methods and position their work within the broader context of machine learning advancements. The research's strengths include its detailed methodology and contribution to generating anti-detection adversarial samples. However, the study lacks in-depth discussion of broader implications, such as ethical concerns and scaling challenges. Exploring alternative adversarial techniques and potential biases could improve robustness.

## III. Research Methodology

### A. Experimental Setup and Dataset Generation

The London Metropolitan University Digital Forensics Laboratory created a comprehensive malware dataset to implement and validate the proposed reinforcement learning malware investigation system [17]. We set up thirteen virtual machines in a secure network to prevent unintended malware spread. We uploaded different ISO files of the Windows operating system to ensure a diverse test environment. We introduced malware to each virtual machine, took snapshots of infected and uninfected states, and produced 26 RAM files. To uncover malware behaviours, we analysed these files using the Volatility System. Finally, we created a detailed workflow diagram to facilitate the training and validation of the model.

### B. Malware Workflow Diagram Creation

The research methodology covers the experimental setup, dataset generation, and development of a malware analysis workflow. This workflow is central to the reinforcement learning System, integrating techniques such as data collection, examination, and analysis [18]. We analysed live memory dumps from *13* versions of Windows, both infected and uninfected, using the Volatility System to detect anomalies and malware artefacts. The analysis phase employs static, signature-based, behavioural techniques, and machine learning algorithms. The workflow diagram maps typical malware behaviours and improves post-incident forensic investigations, supporting the reinforcement learning model's training and validation.

### C. Q-Learning Terminologies

We implement the proposed Reinforcement Learning Post-Incident Malware Investigation Model in the following sections. Key terminologies are briefly defined. The ***Environment*** is the world where the agent operates. The ***Agent*** learns by interacting with the environment. ***States*** (s) represent the agent's position, while an ***Action (a)*** is any move the agent can take, leading to either a reward or penalty. ***Episodes*** signify the end of a stage, either through success or failure. For each state-action pair, the agent manages ***Q-values*** in a ***Q-Table***. ***Temporal Differences (TD)*** compare current and previous state-actions. The ***learning rate*** controls how new information replaces old. A ***policy*** maps states to actions, while the ***Discount Factor*** weighs future rewards. The ***Bellman Equation*** relates Q-values across state-action pairs, and the ***Epsilon-Greedy*** strategy balances *exploration* and *exploitation*.

### *D. The Unified Markov Decision Process*

The Unified Markov Decision Process (MDP), illustrated in ***Figure 1***, brings together various MDP components into a single framework, offering a complete overview. This integration enables the agent to efficiently explore the environment and make well-informed choices in the context of malware analysis.

### *E. Proposed RL Post-Incident Malware Investigation System*

The main components of the Reinforcement Learning Post-Incident Malware Investigation System consist of six key elements, as shown in ***Figure 2***. These elements include data acquisition, workflow diagram mapping, implementation of the MDP model, environmental dependencies, the MDP solver, and continuous learning with adaptation. Initially, data acquisition involves capturing live memory dumps from Windows operating systems. Subsequently, data analysis focuses on examining the acquired data to detect anomalies, indicators of compromise, and potential malware artifacts. The workflow diagram provides a detailed approach for identifying malware infections through static analysis, signature-based techniques, behavioural analytics, and machine learning algorithms. The AWK module carries out feature extraction from identified processes. In addition, listing DLLs is important for monitoring the DLLs loaded by each process. Monitoring open handles is also critical for tracking the open handles associated with each process. Network data collection ensures the capture of all relevant network-related information. We conduct registry hive analysis to identify registry hives and enumerate their keys. We duplicate processes into executable files and compare them with known malware databases to determine their malicious or benign nature. The state spaces are constructed to match the malware workflow diagram, encompassing ***67*** unique states. Actions are defined based on this workflow, ranging from ***three*** to ***ten***, which exposes the agent to ***109*** different actions within a defined environment. As part of our work on the MDP solver, we create unified Markov Decision Process (MDP) models from the proposed RL Post-Incident Malware Investigation Model.

### *F. Algorithm 1 - Implementation of the Q-Learning Algorithm*

**Algorithm 1** implements Q-learning to train an agent for optimal decision-making. It initialises a zero-valued Q-table representing the agent's understanding of the environment, where rows are states and columns are actions. Key parameters like learning rate, discount factor, and exploration probability (initially set to 0.9) are established, along with decay schedules and data storage structures (*storage*, *storage new*, *reward list*). The main loop runs over a specified number of episodes, resetting the environment and variables at each episode's start. Within each episode, a ***greedy policy*** selects actions: if the probability is high, it explores by choosing a random action (***exploration***); otherwise, it exploits by selecting the action with the highest Q-value for the current state (***exploitation***). The environment executes the action, providing the next state, reward, and a done flag indicating the episode's end. Q-values are updated using the Finally, we return the Q-table and storage data, which summarise the learnt policy and training data. The process begins with the initialisation phase, where three custom environments (*env_new1*, *env_new2*, and *env_new3*) are defined using a specified *MDP function*. The algorithm iterates over a list of names (*name_list*), and for each name, it assigns the appropriate environment by configuring the Markov Decision Process (MDP) with specific transition probabilities and rewards.

### *G. Algorithm -2 Iterating Learning Rates Variation over MDP Environments*

**Algorithm 2** is a method that *trains* and *saves* models by utilising various *learning rates (LRs)* across different environments. The algorithm starts with an initialisation phase, where it defines multiple environments (envs) and creates an empty dictionary called '*final dict*' to hold the results. Training parameters are then set, which include a list of learning rates ranging from ***0.001 to 0.9***, along with storage for the corresponding Q-tables, intermediate data, rewards, and other storage structures. For each learning rate (***lr***) in the list of learning rates (***lrs***), the algorithm applies the '***new q learning***' algorithm. Afterward, the results are appended to the '***final dict***', ensuring an organised process for model training and results recording across various environments.

## IV. MDP Models Integration and Implementation

### *A. An Overview of Our Three Proposed MDP Environments*

The **`BlankEnvironment`** represents the proposed Markov Decision Process using the Malware Workflow Diagram, encompassing *states, actions, rewards, transition probabilities*, and the completion status of an *episode*. It consists of a discrete action space with ***10*** possible actions and an observation space comprising ***67*** observations, applying a default step penalty of ***-0.04*** and awarding a reward of ***2*** for detecting malware. The **`BlankEnvironment_with_Rewards`** provides a reward of ***2*** for correctly identifying malware in all terminal states and ***4*** for correct identification at earlier stages, promoting accurate classifications. On the other hand, the **`BlankEnvironment_with_Time`** enforces a stricter penalty of ***-0.01*** per step to encourage timely malware detection by deterring the agent from making excessive moves. In both environments, rewards are treated as hyperparameters, fine-tuned to achieve optimal agent performance.

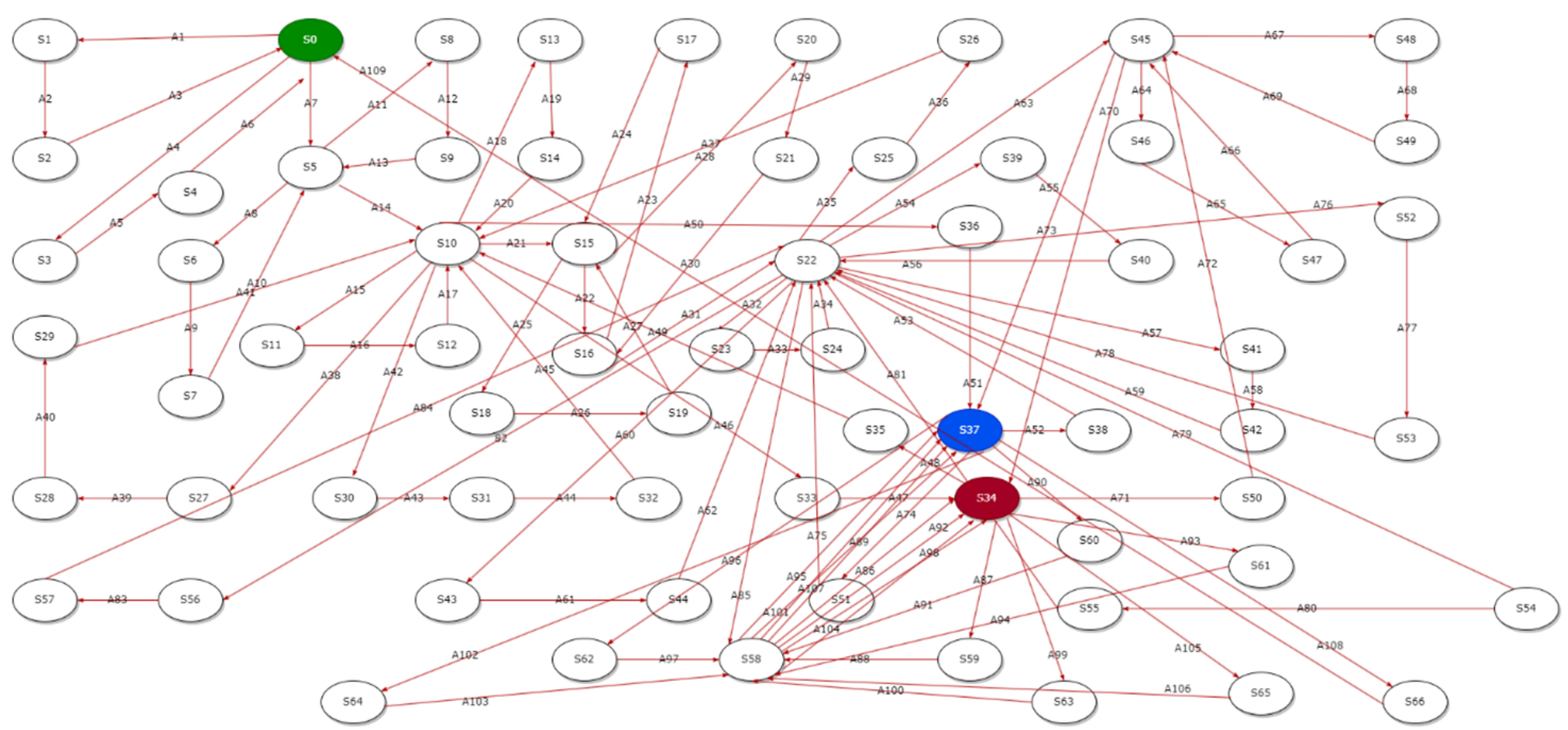

**Fig. 1:** *Overall Markov Decision Process (MDP) Model*

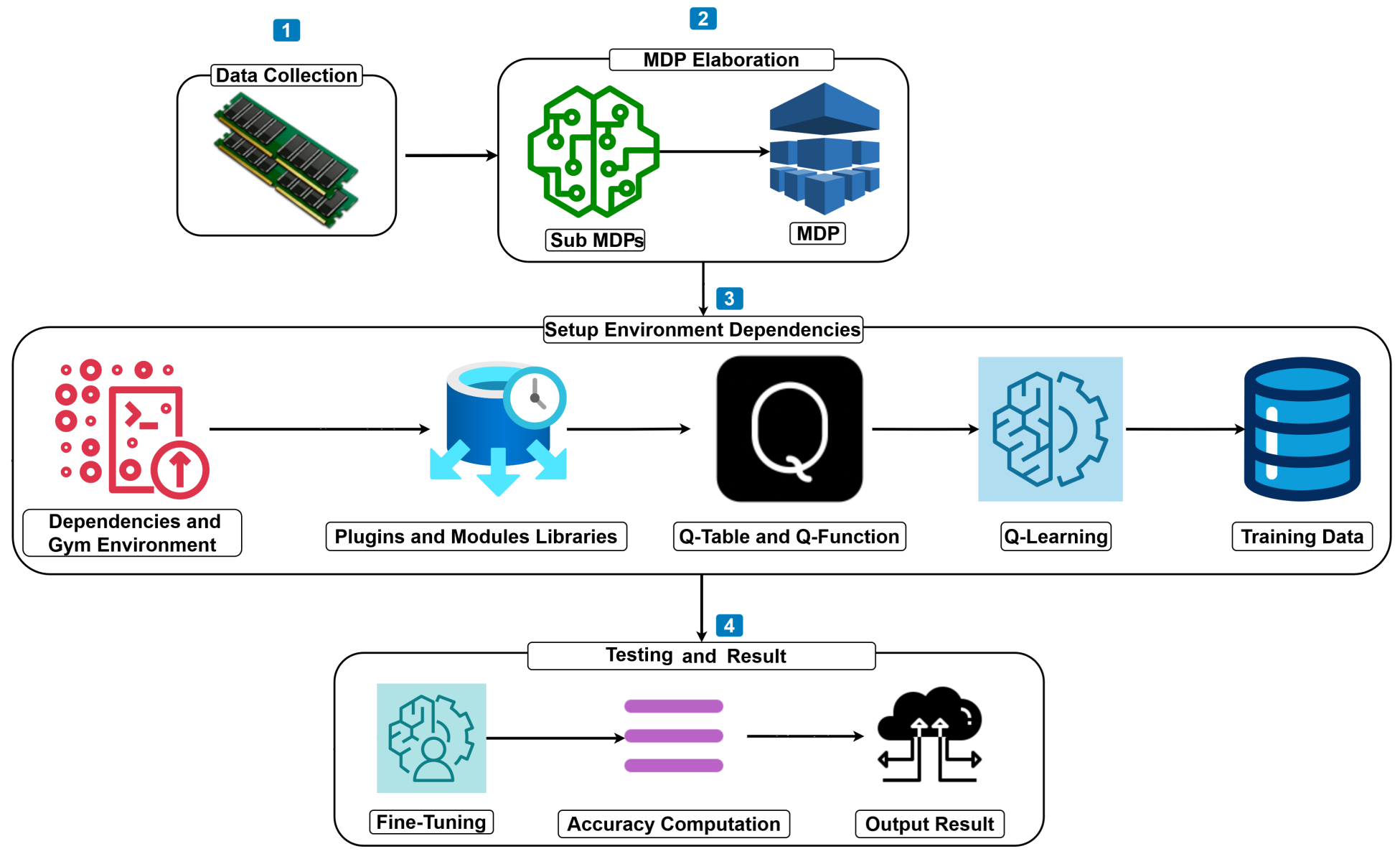

**Fig. 2:** *The Proposed RL Post-Incident Malware Investigation System*

## B. *Implementation of the BlankEnvironment*

In ***Figure 3***, we introduce a new class called '***`BlankEnvironment`***', which extends from '***gym.Env***', signifying its function as a gym-compatible environment. Within the `BlankEnvironment` class, the constructor sets up the instance of the class. A subsequent variable specifies the environment's action and observation spaces. The action space is discrete, containing ***10*** possible actions, while the observation space is also discrete, with ***67*** possible observations. The variable '***self.state = 0***' initialises the environment's starting state as '0'. We also initialise '***self.P = dict()***' as an empty dictionary to store the transition probabilities in the later code sections.

```
class BlankEnvironment(gym.Env):
    def __init__(self):
        # Define action space and observation space
        self.action_space = gym.spaces.Discrete(10)
        self.observation_space = gym.spaces.Discrete(67)

        #self.action_space = np

        self.state = 0

        self.P = dict()
```

**Fig. 3:** *Initialise and Implement the BlankEnvironment Class*

**Algorithm 1** *Q-Learning Implementation*

1: **Initialization**:
- Initialize Q-table with zeros. (defines the state of the agent)
- Set parameters: learning rate ($\alpha$), discount factor ($\gamma$), exploration probability ($\epsilon = 0.9$), and decay schedule.
- Initialize storage structures: storage, storage_new, reward_list.

2: **for episode = 0, 1, . . . , episodes do**
3: Reset environment and variables:
- Reset the environment to obtain the initial state.
- Initialize episodic reward and step counter.
- Store current epsilon value.

4: **while not done, do**
5: Select action using $\epsilon$-greedy policy:
- if $\epsilon <$ rand() then
- action ~ Uniform(noA)
- else
- action = $\max_a Q(\text{state}, a)$

6: **Execute action:**
- Act in the environment and observe the next state, reward, and done flag.
- Update episodic reward.
- Increment steps counter.

7: Update Q-value and Compute the maximum future Q-value for the next state.
8: **Calculate the new Q-value**
9: Update the Q-table with the new Q-value.
10: Store Q-value updates if specific conditions (e.g., state, action) are met.
11: Update state: Set the current state to the next state AND Append the (current_q, new_q, episode, action) to the storage list.
12: Decay $\epsilon$: Reduce epsilon based on the decay schedule.
13: Check convergence: if $|\text{new_q} - \text{current_q}| <$ threshold and new_q $\neq$ current_q then
14: Break the loop
15: Append (episodic_reward, episode, steps) to reward_list.
16: Update $\epsilon$:
- $\epsilon \leftarrow \epsilon - (\epsilon\text{_decay_value} \times 0.5)$

17: Return results:
- **Return Q-table, storage, reward_list, and storage_new.**

18: **end**

### C. Implementation of the BlankEnvironment with Rewards

The ***`BlankEnvironment with Rewards`*** represents a distinct implementation in comparison to the ***`BlankEnvironment`***. In ***`BlankEnvironment with Rewards`***, actions that lead to terminal states receive a reward of *2*, unlike the *-0.04* reward given in the ***`BlankEnvironment`***. The reward function in ***`BlankEnvironment_with_Rewards`*** is adjusted at the end of an episode, as indicated by the done flag. The done flag assigns the fourth element's value to the variable named done, which reflects whether the episode has concluded.

**Algorithm 2** *Iterating Learning Rates Variation over MDP Environments*

1: **Initialization**: Defining different envs and empty final dict
2: **for** name **in** name_list **do**
1) Assigning the right env (MDP): Using diff transition probs and rewards to create the MDP
2) Defining and resetting the training params:
   - Learning rates list
   - outputs, store and rewards dictionaries
3) **for** lr **in** lrs **do**
   a) Performing the new_q_learning algorithm
   b) Storing everything by appending in the final_dict
4) **end for**

3: **end**

When the done flag detects that the done variable is set to True, the reward variable is assigned a positive value of *4*. This modification accounts for the impact of changing the reward at the conclusion of the episode.

### D. Implementation of the BlankEnvironment with Time

In ***`BlankEnvironment_with_Time`***, the agent encounters a harsher negative reward of ***-0.1*** per step, in contrast to the usual penalty of ***-0.04*** present in the other two environments. The design of this approach encourages the agent to swiftly identify malicious files by taking the shortest possible route, thereby avoiding any unnecessary actions. Moreover, if the agent prolongs the episode by taking extra steps, it incurs substantial penalties. Crucially, we treat this incentive as a hyperparameter, enabling continuous adjustments. The expression '*done = k[3]*' assigns the fourth element from the tuple '*k*' to the variable 'done', signifying whether the episode has concluded. If 'done' is 'True', the agent receives a reward of ***+4***; otherwise, it is penalised with ***-0.1***. Subsequently, the tuple 'new k' is generated by retaining the original values from '*k*' but updating the reward. This modified tuple is then returned for further interactions with the environment.

### E. Iterating MDP Environments over Learning Rates

We developed a Python script to iteratively test the three Markov Decision Process (MDP) environments over a range of learning rates (*0.001–0.9*). The `name_list = ['env_new1', 'env_new2', 'env_new3']` contains the names of these environments in a list format. We start by initialising an empty dictionary to store the final outputs and employing a `for` loop to cycle through each environment. Using the *Q-learning function*, we convert the current learning rate to a float and capture the results in dictionaries. We subsequently convert the output into a list and store it within the output dictionary. Finally, we organise the aggregated data into a tuple and place it in the final dictionary, consolidating all results for comprehensive analysis.

## V. Testing and Evaluation

### A. Comparing the Speed of Convergence

We developed a Python script to visualise the convergence rates across three distinct MDP

environments, namely (`env_new1`, `env_new2`, and `env_new3`), which correspond to `BlankEnvironment`, `BlankEnvironment_with_Rewards()`, and `BlankEnvironment_with_Time()`, respectively. Each environment dictionary links learning rates with the number of episodes required for convergence. The code line `x = [float(key) for key in env_new1.keys()]` extracts learning rates as floating-point values from `env_new1`. The command `plt.figure(figsize=(10, 6))` initialises a 10x6-inch plot where scatter plots are created for each environment, with unique colours (blue, red, and green) for differentiation, and lines are added to indicate convergence trends. ***Figure 4*** demonstrates that `BlankEnvironment_with_Rewards` (`env_new2`) achieves the smoothest and quickest convergence. On the other hand, `BlankEnvironment_with_Time` (`env_new3`) exhibits slower convergence due to an increased negative reward function, necessitating higher learning rates and additional computational time. `BlankEnvironment` (`env_new1`) also performs well but experiences slower convergence due to learning rate fluctuations. `BlankEnvironment_with_Rewards` proves to be the most efficient MDP environment with optimal performance at a learning rate of 0.4.

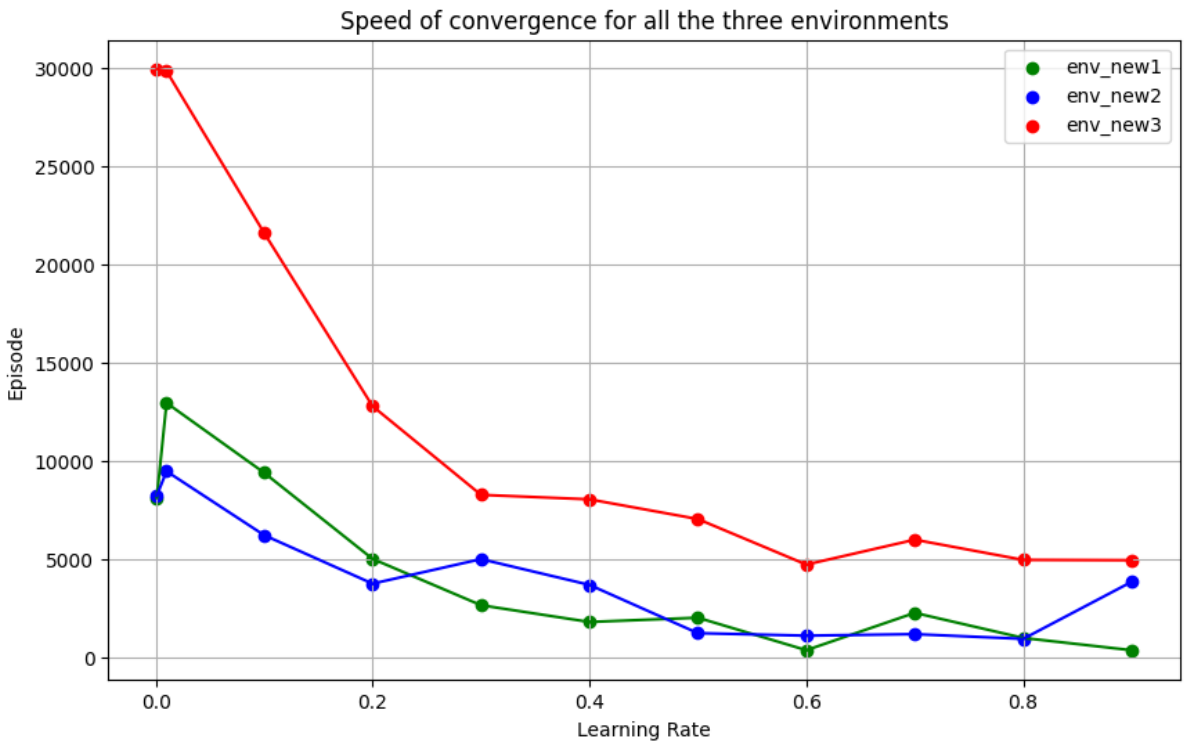

**Fig. 4:** The speed of convergence across the three MDP environments

### B. Command Definitions for State-Based Actions

To facilitate spawning new processes and controlling their input/output/error streams, as well as retrieving their return codes, we incorporated the *`Subprocess`* module in our Python script. Next, we created and filled an empty dictionary, *`my_dict`*, using **key-value** pairs. Each key corresponds to a distinct ***state***, and its value is a list of associated ***commands***. For instance, *state 0* contains a command for cloning a GitHub repository, whereas *state 10* comprises commands related to Windows system information and the registry. Furthermore, *states 15 through 45* hold various commands, some of which include special characters and options such as `-pid` and `-o`.

## VI. Results and Discussion

### A. The Agent Decision-Making Processes

We developed a Python script to initialise two lists, *`ideal_list`* and *`pred_list`*, each populated with integer values representing ***actions*** for particular ***states*** within our reinforcement learning MDP framework. Specifically, *`ideal_list`* stores the optimal actions for states ***0 through 66***, while *`pred_list`* holds the corresponding predicted actions for these states. For instance, *state 0* has an ideal *action* of *0* compared to a predicted *action* of *2*. Each index in both lists aligns with a distinct state, enabling a direct comparison between predicted actions and ideal actions. This comparison is essential for evaluating model performance across the three environments, which differ by applied learning rates.

### B. Python Function to Evaluate Predictive Model Accuracy

To assess the accuracy between predicted and ideal actions, we developed a Python function called *`get_acc`*. The function begins by initialising two counters, *`true`* and *`false`*, both set to zero, to track correct and incorrect predictions, respectively. Using the *`zip()`* method, the function concurrently traverses both *`ideal_list`* and *`pred_list`*, comparing each corresponding pair of elements. If a match is found, *`true`* is incremented; otherwise, *`false`* increases. After completing the iteration, the accuracy is calculated by dividing *`true`* by the total number of comparisons (*`true + false`*), with the result formatted to five decimal places. Finally, the ***accuracy*** is displayed. This function provides a practical tool for evaluating prediction accuracy in reinforcement learning applications, demonstrating an accuracy rate of 94% upon execution.

### C. *`get_acc`* Function for Accuracy Computation

The implemented Python function *`get_acc`*, processes three different environments, namely `env1`, `env2`, and `env3`. Each environment is represented by dictionaries (`q1_dict`, `q2_dict`, and `q3_dict`), storing relevant data points. For these environments, we define sets of $x$ and $y$ coordinates: `env1_x` and `env1_y`, `env2_x` and `env2_y`, as well as `env3_x` and `env3_y`, to represent their unique data distributions. Subsequently, the function generates three scatter plot traces using `go.Scatter`, where it specifies data points, display modes (lines and markers), labels, and marker colours for each environment. Following this, a layout is configured for the plot to include an informative title, x-axis and y-axis labels, and a hover mode to enhance interactivity. A figure object is then constructed by combining the layout and traces, allowing for comprehensive visualization. Finally, the plot is rendered with `fig.show()`, effectively showcasing the accuracy calculation for various learning rates across *three MDP environments*, as illustrated in ***Figure 5***. This visualisation indicates that among the environments, ***env2***, with a ***learning rate of 0.4***, achieves the highest performance.

### D. Plotting the Proposed Model Command Execution Timings

As a consequence of tracking state transitions through our proposed reinforcement learning model for post-incident malware investigation, we generate a trajectory based on selected actions and resulting states, which drive a sequence of environmental state changes. Using the execution timings available in the Google Collaborative Environment, we visualised the command execution timings of our proposed model. For this purpose, we developed a new dictionary, `command_timings_dict`, to store key-value pairs that link states to their respective action trajectories. Next, we implemented Python code to create a multi-plot figure using Plotly, designed to analyse execution times of various malware commands such as WannaCry, Cerber, and Cridex. This code initialises a figure containing three vertically arranged subplots, each titled and spaced to enhance readability. We then add individual line plots for WannaCry, Cerber, and Cridex commands to the first, second, and third subplots, respectively, ensuring each is visually distinct with unique colours and markers. We adjust the figure's layout to set its dimensions and align the title in the center. We further customise each subplot's X-axis labels and assign the y-axes the label 'Time (seconds)'. Finally, the completed graph is displayed using `fig.show()`, as illustrated below in ***Figure 6***.

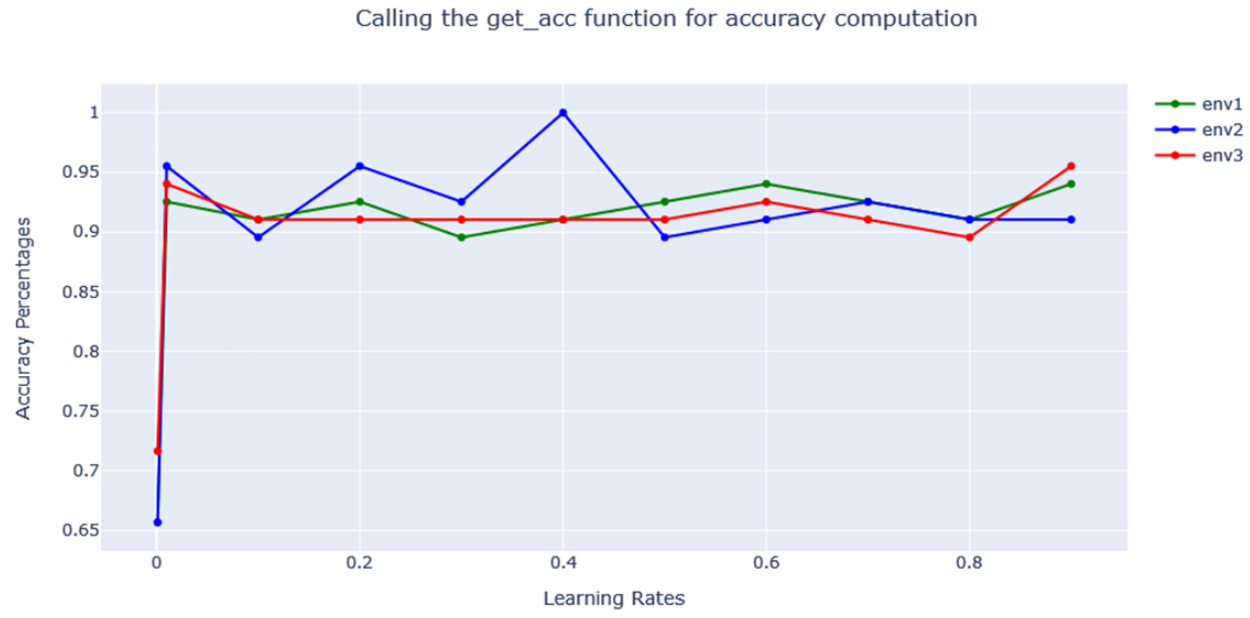

**Fig. 5:** *Calling the* `get_acc` *function for accuracy computation*

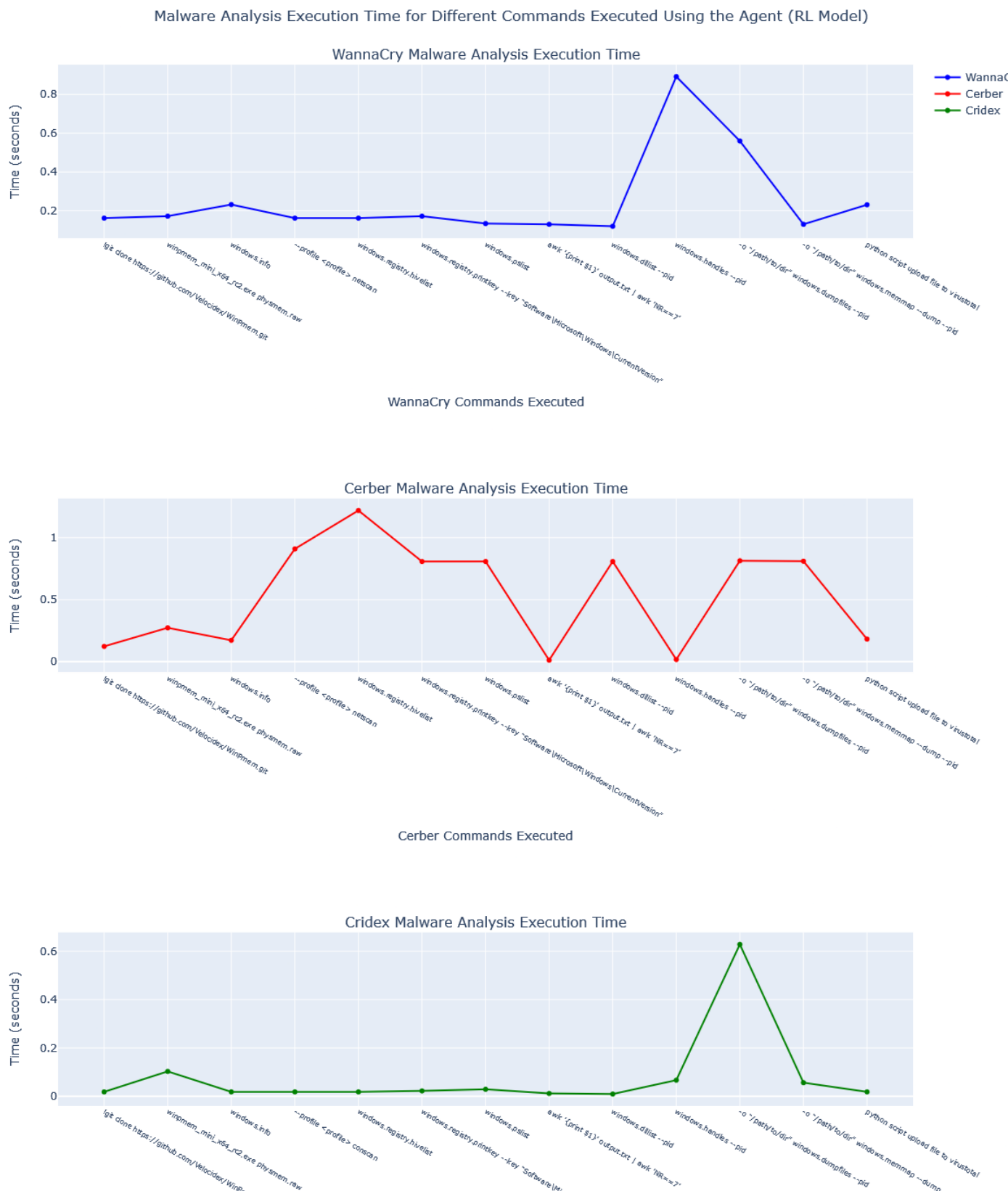

**Fig. 6:** *Malware Analysis Execution Time for Different Commands Executed Using the Agent (RL Model)*

### E. RL Model Evaluation Against Malware Detection

The proposed reinforcement learning (RL) model demonstrates notable advantages over traditional methods, such as signature-based and heuristic approaches, which often falter in identifying new or polymorphic threats due to their reliance on predefined patterns. In contrast to these conventional techniques that require regular updates, our proposed RL dynamically adapts by continuously learning and refining its detection capabilities through interaction with the environment, leveraging methods like Q-learning and Markov Decision Processes. This dynamic adaptability significantly reduces both false positives and false negatives, thereby enhancing detection accuracy, with our RL-based models achieving up to a 94% improvement in identifying adversarial malware. Additionally, the Reinforcement Learning self-optimisation framework offers a robust defence against zero-day threats, providing a more adaptive response than static systems, which are constrained by their limited adaptability to novel or obfuscated malware. In addition, to ensure scalability, different reward structures across Markov Decision Processes (MDP) environments prioritise efficient decision-making and faster convergence, making our model well-suited for resource-constrained scenarios. Thus, the RL model not only offers a more robust solution for malware forensics but also integrates seamlessly with existing cybersecurity frameworks, paving the way for future enhancements through hybrid approaches and advanced optimization techniques. Consequently, the integration of RL into malware forensics represents a transformative step toward smarter and more resilient cybersecurity solutions.

### F. Research Findings and Recommendations

As a result of automating forensic tasks, particularly the identification of malware artefacts through a structured RL workflow based on Q-learning, the model demonstrates effective capabilities in identifying and classifying malware. However, the model's accuracy is contingent upon the diversity of malware samples, necessitating broader datasets to enhance its performance. To address computational constraints, optimizing RL models for efficiency is essential, especially within resource-limited environments. The use of parallel processing techniques and cloud-based infrastructure can enable the model to manage larger datasets and a wider variety of malware types, thus enhancing scalability. This adaptability allows the model to dynamically scale with increasing data volumes, making it particularly valuable for real-time incident response. Techniques such as prioritising state transitions based on malware severity or employing approximate Q-learning methods can further contribute to computational efficiency, reducing training times. Additionally, model compression methods such as pruning and quantization can decrease memory usage and computational demands, facilitating deployment in environ-ments with limited resources. These collective enhancements substantially improve the RL model's applicability to large-scale, real-world cybersecurity operations. We recommend integrating this model within existing security infrastructures to leverage RL's adaptive learning alongside traditional forensic processes, thereby creating a more resilient defence mechanism.

## VII. Conclusion

This paper presents a novel reinforcement learning (RL) model and system for post-incident malware forensics investigations, designed to surpass traditional capabilities and enhance the efficiency of malware analysis. The model expedites the identification of both known and unknown threats by incorporating advanced techniques, such as live memory dumps from Windows systems. The Proposed unified Markov Decision Process (MDP) was developed, featuring three distinct environments with each designed with unique reward mechanisms to optimise malware analysis. The experiments used simulations with malware such as WannaCry, Cridex, and Cerber to demonstrate the impact of learning rates and environmental complexity on the model's performance. This shows how important hyperparameter tuning and continuous learning are for improving the performance of RL models. To further enhance the model, future work should focus on refining the reward functions used in RL to better differentiate between benign and malicious behaviors. In line with these improvements, we propose incorporating a rule-based expert system (RBES) in future work to capture and generalize expertise generated by our system. RBES would make it possible to directly apply learned information to future investigations, especially when working with similar machine architectures and configurations. This would make the model more consistent and useful for re-investigations involving similar platforms, such as targeted systems with similar builds. These enhancements promise to bolster the RL model's applicability in real-world cybersecurity operations, contributing to more consistent and effective forensic investigations [20].